\documentclass[prl,twocolumn,nofootinbib, preprintnumbers, superscriptaddress]{revtex4}

\usepackage{amsmath,amssymb,slashed,braket,bm}
\usepackage{graphicx}
\usepackage{epstopdf}
\usepackage{float}
\usepackage[colorlinks=true,
            linkcolor=blue,
            urlcolor=blue,
            citecolor=green,          
            bookmarks=true,
            bookmarksnumbered=true,
            breaklinks=true,
            pdfpagemode=Fullscreen,
            pdfstartview=FitBH]{hyperref}

\usepackage[normalem]{ulem}
\usepackage{color}

\definecolor{Orange}{cmyk}{0,0.61,0.87,0}
\definecolor{JungleGreen}{cmyk}{0.99,0,0.52,0}
\definecolor{OliveGreen}{cmyk}{0.64,0,0.95,0.40}
\definecolor{Brown}{cmyk}{0,0.81,1,0.60}
\definecolor{RoyalBlue}{cmyk}{0.71,0.53,0,0.12}
\definecolor{Gray}{cmyk}{0,0,0,0.40}
\definecolor{LightPink}{cmyk}{0.0,0.25,0,0}
\definecolor{LLightPink}{cmyk}{0.0,0.10,0,0}
\definecolor{LightBlue}{cmyk}{0.25,0,0,0}
\definecolor{LightGray}{cmyk}{0,0,0,0.2}

\usepackage{xcolor}
\definecolor{gesfpurple}{rgb}{0.47,0.19,0.42}

\definecolor{gesflanse}{rgb}{0.00,0.50,0.50}

\definecolor{gesfblue}{rgb}{0.08,0.42,0.76}

\definecolor{gesfred}{rgb}{1,0,0}

\definecolor{gesfwhite}{rgb}{1,1,1}

\definecolor{gesfblack}{rgb}{0,0,0}

\definecolor{Plum}{rgb}{0.56, 0.27, 0.52}

\newcommand{\geqn}[1]{Eq.\,\hypersetup{linkcolor=blue}(\ref{#1})\hypersetup{linkcolor=blue}}
\newcommand{\gfig}[1]{{\hypersetup{linkcolor=violet}Fig.\,\ref{#1}\hypersetup{linkcolor=blue}}}

\graphicspath{{figs/}}

\begin{document}

% \title{\Large Probe Light Dark Matter by Cosmic Gravitational Focusing}
\title{\Large Probing Light Dark Matter with Cosmic Gravitational Focusing}

\author{Shao-Feng Ge}
\email[Corresponding Author: ]{gesf@sjtu.edu.cn}
\affiliation{State Key Laboratory of Dark Matter Physics, Tsung-Dao Lee Institute \& School of Physics and Astronomy, Shanghai Jiao Tong University, China}
\affiliation{Key Laboratory for Particle Astrophysics and Cosmology (MOE) \& Shanghai Key Laboratory for Particle Physics and Cosmology, Shanghai Jiao Tong University, Shanghai 200240, China}

\author{Liang Tan}
\email[Corresponding Author: ]{tanliang@sjtu.edu.cn}
\affiliation{State Key Laboratory of Dark Matter Physics, Tsung-Dao Lee Institute \& School of Physics and Astronomy, Shanghai Jiao Tong University, China}
\affiliation{Key Laboratory for Particle Astrophysics and Cosmology (MOE) \& Shanghai Key Laboratory for Particle Physics and Cosmology, Shanghai Jiao Tong University, Shanghai 200240, China}

\begin{abstract}

We investigate the possibility of using the cosmic
gravitational focusing (CGF) to probe the minor light dark
matter (DM) component whose mass is in the range of
$(0.1 \sim 100)$\,eV.
Being a purely gravitational effect, the CGF offers
a mode-independent probe
that is complementary to the existing ways such as
Lyman-$\alpha$ and $\Delta N_{\rm eff}$.
Such effect finally leads to a
dipole density distribution that would affect the galaxy
formation and hence can be reconstructed with galaxy
surveys such as DESI. Both the free-streaming and
clustering limits have been studied with analytical formulas
while the region in between is bridged with interpolation.
We show the projected sensitivity at DESI with
the typical phase space distribution of a freeze-in
DM scenario as illustration.

\end{abstract}

\maketitle 

{\bf Introduction} -- 
The DM plays very important roles
in the evolution and structure formation of our Universe
\cite{Bertone:2016nfn,Young:2016ala,Arbey:2021gdg}.
For redshift $z \gtrsim 10$ when the dark energy
has not started to dominate, the behavior and history of
our Universe is mainly determined by radiation
and matter. Of the later, more than 80\% are
contributed by DM while the remaining by the
ordinary matter. To be exact, DM is
more than 5 times of the ordinary matter.
It is then a fair question to ask whether the
DM has just a single type or actually possesses
multiple components. Since the ordinary matter
world is already a combination
of various isotopes that are formed by at
least three building blocks (proton, neutron,
and electron), it is natural for the DM sector
to also have several species.

The particle physics provides various DM candidates
\cite{Bertone:2004pz,Feng:2010gw}
that are not just conceptually neat with unified
quantum field theory description in the same way
as the ordinary matter but also very
simple with typically just mass
and coupling strength as the only parameters to
explain the observed DM phenomena from both astronomy
and cosmology. Of them, the particular interesting ones
include the Weakly Interacting Massive Particles (WIMP)
\cite{Steigman:1984ac,Goodman:1984dc,Arcadi:2017kky,Roszkowski:2017nbc,Arcadi:2024ukq}
that participate the weak gauge interactions with
mass typically at GeV$\sim$TeV, the sterile neutrino at keV scale
\cite{Dodelson:1993je,Shi:1998km,Kusenko:2009up,Shakya:2015xnx,Drewes:2016upu,
Abazajian:2017tcc,Boyarsky:2018tvu, Kopp:2021jlk,Ivanchik:2024mqq}
suggested by the observed anomalies in
neutrino oscillation experiments and astrophysical
observations, and the axion
\cite{Preskill:1982cy,Marsh:2015xka,Adams:2022pbo}
motivated by the strong CP problem with even lighter mass. 
The mass spectrum of particle DM candidates spans
around 100 orders
from the smallest fuzzy DM at $10^{-22}\,{\rm eV}$ to
the astrophysical primordial black holes (PBH) with
masses of $10^{50}$\,g ($\sim 10^{88}$\,eV) \cite{Carr:2020gox}.
It is possible for the DM candidates to have totally
different masses.

Besides mass, another important property is
whether the DM is cold, warm, or even hot \cite{Rubakov:2017xzr}.
It can have significant
effect on the large scale structure (LSS) of our Universe today.
The comparison between the galaxy survey and the theoretical
N-body simulation shows strong preference of the cold DM (CDM) 
\cite{Blumenthal:1984bp,Liddle:1993fq,Ostriker:1993fr,Kurek:2007tb}.
Such conclusion is sometimes strengthened to a claim that DM has to be
cold. Nevertheless, this is based on the assumption
that there is just one DM type. If the DM sector
has multiple components, it is perfectly fine to have
CDM as the major component with some warm DM (WDM)
as a minor contribution. Especially, a mixture of
CDM and WDM can help solving the small scale problem
\cite{Harada:2014lma,Kamada:2016vsc}.

The small-scale effect of WDM can be probed by
Lyman-$\alpha$ \cite{Viel:2013fqw,Murgia:2018now,Irsic:2023equ},
Milk Way satellite galaxies \cite{DES:2020fxi},
weak lensing \cite{Inoue:2014jka},
strong lensing \cite{Gilman:2019nap,Zelko:2022tgf,Keeley:2024brx},
galaxy UV luminosity function 
\cite{Dayal:2023nwi,Liu:2024edl,Padmanabhan:2024vuc},
and
stellar streams \cite{Banik:2018pjp,Banik:2019smi}.
If DM is a fermion, it should also be subject to
the fermion degenerate gas (Tremaine-Gunn)
constraints at the galaxy \cite{Tremaine:1979we} and
cosmological \cite{Carena:2021bqm} levels.
These observations require the WDM mass 
$ m_{\rm WDM} \gtrsim \mathcal O(1)$\,keV.
Therefore, the light WDM with mass below keV can
contribute only a fraction of the total DM.
The mixed DM scenario with both CDM and WDM,
has been widely explored
\cite{Masiero:1994bx,Boyanovsky:2007ba,Boyarsky:2008xj,
Anderhalden:2012qt,Harada:2014lma,Lello:2015uma,
Kamada:2016vsc,Parimbelli:2021mtp,
Giri:2022nxq,Keeley:2023ive,Horner:2023cmc,Peters:2023asu,
Inoue:2023muc,Euclid:2024pwi,Tan:2024cek,Garcia-Gallego:2025kiw,
Verdiani:2025jcf,Tadepalli:2025gzf,Celik:2025wkt}
with various models
\cite{Jedamzik:2005sx,Baer:2008yd,Baer:2010kw,Ibe:2013pua,Borah:2017hgt}.
The current Lyman-$\alpha$ gives a constraint on the
WDM fraction $F_{\rm WDM} < \mathcal O (0.1)$
for the light DM mass $m_{\rm WDM} < 100\,{\rm eV}$
\cite{Hooper:2022byl}.
If the WDM mass below $m_{\rm WDM} < \mathcal O (1)\,{\rm eV}$,
it remains nearly relativistic around recombination
and consequently it will also be subject to the constraint 
on the effective degree of freedom ($\Delta N_{\rm eff}$) 
at the cosmic microwave  background (CMB)
\cite{Brust:2013ova}. The light species has also
been named as hot DM such as
\cite{Davis:1992ui,
Taylor:1992zh,
Liddle:1993ez,
Bonometto:1993fx,
Masiero:1993zy,
Starkman:1993ik,
Cen:1993ay,
Pogosian:1994ns,
Masiero:1994bx,
Pogosyan:1995pkv,
Pogosian:1994tx,
Gorski:1995ys,
Borgani:1995rq,
Caldwell:1995iw,
Kofman:1995ds,
Strickland:1995bh,
Larsen:1995qa,
Borgani:1996ag,
Hu:1997vi,
Valdarnini:1998zy,
Novosyadlyj:1998bw},
before the neutrino oscillation
was established in 1998.

In this letter, we explore the possibility of using
the cosmic gravitational focusing (CGF) to probe
the minor light DM component. Similar as the cosmic neutrinos
\cite{Zhu:2013tma,Inman:2016prk,Okoli:2016vmd,Nascimento:2023ezc,Ge:2023nnh,Ge:2024kac},
a light DM $X$ can also develop a relative bulk
velocity $\bm v_{Xc} \equiv \bm v_X - \bm v_c$
with respect to the major CDM ($c$).
Then the light DM fluid can be focused
when passing by the CDM halo and develop a density
dipole that can be traced and reconstructed through the
cross correlation between galaxies of different types
\cite{Zhu:2013tma, Ge:2023nnh}.
Being a purely gravitational effect, the CGF effect 
can serve as a model-independent method for
for probing the light DM.

\vspace{2mm}
{\bf Cosmic Gravitational Focusing and Rough Sensitivity Estimation} -- 
As studied earlier, the CGF would
lead to higher density in the downstream of a light particle
fluid such as the cosmic neutrinos
\cite{Zhu:2013tma, Okoli:2016vmd, Ge:2023nnh,Ge:2024kac}.
After substracting the average density, the remaining
overdensity $\delta(\bm x)$ mainly behaves like a dipole,
$\delta(\bm x) = - \delta(- \bm x)$. With Fourier transformation,
$\tilde \delta(\bm k) \equiv \int \delta(\bm x) e^{- i \bm k \cdot \bm x} d^3 \bm x$,
such density dipole becomes an imaginary
contribution, $\tilde \delta^*(\bm k) = - \tilde \delta(\bm k)$,
in the wave-number ($\bm k$) space \cite{2009JCAP...11..026M}.
Then, the total matter overdensity $\tilde \delta_m$
contains the major CDM and the minor light DM
contributions as real and imaginary parts,
$\tilde \delta_m \rightarrow (1 + i \tilde \phi_X) \tilde \delta_m$,
respectively. Below the free-streaming scale $k_{\rm fs}^{-1}$ of
the light DM $X$, $|\bm k|^{-1} < k_{\rm fs}^{-1}$,
the imaginary phase $\tilde \phi_X$ for a thermal relic
\cite{Ge:2023nnh} is,
\begin{align}
  \tilde \phi_X
\equiv
  \frac{G a^2}{|\bm k|^2}
  ({\bm v}_{X c} \cdot \hat {\bm k})
\left( 
    m_X^4 f_0
  + 3 m_X^2 T_A^2  f_1
  + 2 T_A^4 f_2
\right),
\label{eq:phiRel_main}
\end{align}
where $\hat{\bm k}$ is the unit vector of
the wave number $\bm k$.
In addition, $G$, $a$, and $T_A$ are the Newton constant, 
scale factor, and spectrum parameter
that controls the light DM momentum distribution, respectively.
Note that the spectrum parameter $T_A$
redshifts in the same way as temperature,
$T_A (a) = T_{A0} / a$ where $T_{A0}$ is the value
today, which would
be further discussed in \geqn{eq:fp_freeze_in}.
The coefficients
$f_n (y_i) \equiv  g_X  \int_{y_i}^\infty dy y^{2n} d f_X(y) / d y $
are obtained from the integration of the phase space distribution
$f_X (\bm p)$
with $y \equiv |\bm p| / T_A$ and the lower limit
$y_i \equiv m_X | \bm v_{X c} \cdot \hat{\bm k} |/ T_A$
\cite{Ge:2023nnh}. 
With the light DM $X$ being non-relativistic
($m_X \gg 2.7$\,K $\sim 10^{-4}$\,eV) today,
the first term dominates the $m^4_X$ dependence. 
So the CGF effect becomes stronger for a heavier DM $X$.

As elaborated in previous studies
\cite{Zhu:2013tma, Okoli:2016vmd, Ge:2023nnh,Ge:2024kac},
the cosmic relic neutrino with sub-eV mass can already
have sizable effect.
Comparing with the existing LSS and CMB constraints
\cite{Ge:2023nnh,Ge:2024kac}, the CGF effect can give
at least similar sensitivity on the neutrino mass.
While the cosmic relic neutrinos contribute 0.3\% of the
total energy of our Universe today, the CGF constraint
on the light DM fraction $F_X$ of the total DM density
should reach $0.3\% / 27\% \approx 1\%$ if the light DM
has roughly the same mass as neutrinos. Not to say
the light DM can have a much larger mass. With
$m^4_X$ dependence in \geqn{eq:phiRel_main},
the sensitivity can significantly enhance for eV mass
to easily exceed the existing constraints.

\vspace{2mm}
%{\bf Freeze-in DM Phase Space Distribution} -- 
{\bf Freeze-in DM with Modified Mass Scaling Behavior} -- 
The thermally produced light DM with a mass below $1\,{\rm MeV}$
is almost excluded by the big-bang nucleosynthesis,
since it would contribute too much $\Delta N_{\rm eff}$
\cite{Sabti:2019mhn,Sabti:2021reh,Giovanetti:2021izc,Chu:2022xuh}. 
Hence, the light DM is mostly generated from
the freeze-in mechanism
\cite{Hall:2009bx,Bernal:2017kxu,Dvorkin:2020xga},
such as the two-body decay
\cite{Heeck:2017xbu, Boulebnane:2017fxw, Kamada:2019kpe},
or two-to-two processes \cite{DEramo:2020gpr}.
A typical phase space distribution of such light DM $X$
\cite{Heeck:2017xbu, Boulebnane:2017fxw, Kamada:2019kpe,DEramo:2020gpr,Huang:2023jxb} is
\begin{align}
  f_X(\bm p) 
\approx 
  C_X \frac{e^{- |\bm p| / T_A (a)}}{\sqrt{|\bm p| / T_A (a)}},
\label{eq:fp_freeze_in}
\end{align} 
where $\bm p$ is the DM $X$ momentum, and
$T_A(a) \equiv T_{A0} / a$ is a spectrum parameter inherited from
the freeze-in process. We take $T_{A0} = 10^{-4}$\,eV
as a characteristic value today. The normalization
coefficient $C_X$
\cite{Heeck:2017xbu,Boulebnane:2017fxw,Kamada:2019kpe,DEramo:2020gpr,Huang:2023jxb}
can be parameterized by its current energy density,
\begin{align} 
  \rho_{X0} 
\equiv
  g_X m_X 
  \int \frac{d^3 \bm p_0}{(2 \pi)^3}   
  f_{X0}(\bm p_0)
=
  \frac{3 g_X C_X}{8 \pi^{3/2}}  
  m_X  T_{A0}^3,
\label{eq:rho_chi}
\end{align}
where $g_X$ is the number of degree of freedom
for the light DM $X$. The subscript $0$
is for quantities nowadays.

With the light DM phase space distribution $f_X(\bm p)$
in \geqn{eq:fp_freeze_in}, 
those coefficients $f_n$ in \geqn{eq:phiRel_main}
can be integrated analytically,
\begin{align} 
  f_n
=
  -  g_X  C_X
  \left[  
    \frac 1 2
    \Gamma\left( -\frac 1 2 + 2n , y_i \right) 
  + \Gamma\left(\frac 1 2 + 2n , y_i \right)
  \right], 
\nonumber
\end{align} 
where the $\Gamma(x,y)$ is the {\it Upper Incomplete Gamma Function}.
With $m_X \gg T_{A0}$, the first term of \geqn{eq:phiRel_main}
and hence $f_0$ dominate. Using the result
$f_0 = - g_X C_X e^{- y_i} / 2 \sqrt{y_i}$,
and replacing $g_X C_X$ with the current DM
energy density $\rho_{X0}$ in \geqn{eq:rho_chi},
the imaginary phase $\tilde \phi_X$ for a freeze-in
light DM becomes,
\begin{align}
  \tilde \phi_X
\approx 
- \frac{4 \pi^{3/2}}{ 3 }
  \frac{G a^2}{| \bm k|^2}
  ( \bm v_{X c} \cdot \hat{\bm k} )
  \rho_{X0}
  \left( \frac{m_X}{T_{A0}} \right)^3
  \frac{e^{ - y_i }}{\sqrt{ y_i } }.
\label{eq:phichi1_main}
\end{align}
If the DM $X$ is very cold,
it is expected to fully follow the CDM evolution.
In this case, there is no relative velocity,
and no CGF effect at all. Mathematically,
this feature manifests in the last term of \geqn{eq:phichi1_main}
$e^{- \langle y_i \rangle} / \sqrt{y_i} \rightarrow 0$ 
since 
$ \langle y_i \rangle = m_X \langle | \bm v_{X c} \cdot \hat{\bm k} | \rangle / T_A \gg 1$
with $m_X \gg T_{A0}$.

With the freeze-in phase space distribution $f_X(\bm p)$,
the mass dependence of $\tilde \phi_X$ is different
from the previous $m^4_X$ in \geqn{eq:phiRel_main}.
This occurs because the current DM density
$\rho_{X0} \propto m_X$ has absorbed one power of $m_X$.
Additionally, both $y_i$ and
the relative velocity $\bm v_{X c}$
depend on the light DM mass $m_X$.
Putting $y_i = m_X |\bm v_{X c} \cdot \hat{\bm k} | / T_A$ back into
\geqn{eq:phichi1_main}, 
we obtain the mass scaling behavior 
$\tilde \phi_X \propto  |\bm v_{X c}|^{1/2} m_X^{5/2}$
instead of the original $|\bm v_{Xc}| m^4_X$.
Considering the fact that the relative velocity
roughly scales inversely with mass, $\bm v_{Xc} \propto 1 / m_X$,
the mass dependence reduces from the original
$\tilde \phi_X \propto m^3_X$ to $m^2_X$ now.
Since a neutrino mass sum $\sum m_\nu \approx 0.1\,{\rm eV}$
corresponds to the relative size of cosmic neutrinos to the CDM as 
$F_\nu \equiv \Omega_\nu / \Omega_{\rm CDM} \approx 10^{-2} (\sum m_\nu / 0.1\,{\rm eV})$
where $\Omega_\nu$ and $\Omega_{\rm CDM}$ are the
neutrino and CDM energy density fractions of our
Universe today,
a light DM mass $m_X = 1$\,eV is expected to receive
a constraint roughly around
$F_X \equiv \Omega_X / (\Omega_{\rm CDM} + \Omega_X)
\approx \Omega_X / \Omega_{\rm CDM} \lesssim 10^{-4}$.

\vspace{2mm}
% {\bf CGF of light DM} -- 
{\bf Free-Streaming and Clustering Limits} -- 
Note that \geqn{eq:phichi1_main} is valid 
below the free-streaming scale, 
$k_{\rm fs}^{-1}
\equiv  2 \pi \sqrt{2 / 3} \langle |\bm p_X| \rangle / m_X H_0
\approx 0.384 (10\,{\rm eV} / m_X)\,{\rm Mpc}/ h $
\cite{Lesgourgues:2013sjj}
with the average $\langle |\bm p_X| / T_A \rangle = 2.5$
from the phase space distribution in \geqn{eq:fp_freeze_in}.
However, the free-streaming scale of a freeze-in DM 
with $\mathcal O(10)\,{\rm eV}$ mass
is much smaller than 1\,Mpc/$h$ which is
the typical scale of the observed matter power spectrum
\cite{Planck:2018nkj}. So we need to go beyond the
solution \geqn{eq:phichi1_main} for the free-streaming limit.

From the cosmic linear response theory, we can solve the
DM overdensity in the rest frame of CDM \cite{Chen:2020kxi},
\begin{align} 
  \tilde \delta_X
& =
  i \frac{m_X^2 a^2}{\rho_{X 0} }
  \int \frac{d^3 \bm p}{ (2 \pi)^3 }
  \bm k \cdot \nabla_{\bm p} f_{X, {\rm CDM}} (\bm p)
\label{eq:delta_X_nf}
\\
  & \times
  \int_{s_i}^s ds'
  a^2(s') \tilde \Psi (s')
  \exp 
    \left[ 
      - i a \frac{\bm p \cdot \bm k}{m_X} (s - s') 
    \right],
\notag
\end{align}
where $\tilde \Psi$ and $s \equiv \int d t / a^2$
are the gravitational potential and superconformal time.
In addition, the momentum $\bm p$ is defined at
the superconformal times $s$.
The phase space distribution $f_{X, {\rm CDM}} (\bm p)$
for the light DM $X$ is defined in the rest frame of CDM
and is related to its counterpart \geqn{eq:fp_freeze_in}
in the light DM rest frame itself,
$f_{X, {\rm CDM}} (\bm p) \equiv f_{X} (\bm p - m_X  \bm v_{X c})$.
This can be achieved by
changing the integration variable 
$\bm p$ as $\bm p + m_X \bm v_{X c}$
which leads to an extra phase factor
$e^{- i a \bm v_{X c} \cdot \bm k (s - s')}$,
\begin{align} 
  \tilde \delta_X
& =
  i \frac{ m_X^2 a^2}{\rho_{X 0} }
  \int \frac{d^3 \bm p}{ (2 \pi)^3 }
  \bm k \cdot \nabla_{\bm p} f_{X} (\bm p)
  \int_{s_i}^s ds'
  a^2(s') \tilde \Psi (s')
\notag
\\
  & \times
  \exp 
    \left[ 
      - i a \frac{ \bm p \cdot \bm k}{m_X} (s - s') 
    \right]
  e^{- i a \bm v_{X c} \cdot \bm k (s - s')}.
\label{eq:delta_X_nf_fx}
\end{align}

In the clustering limit, $|\bm k|^{-1} \gg k_{\rm fs}^{-1}$
\cite{Chen:2020kxi},
we can expand the first phase factor to the linear order 
$ \exp \left[ - i a \bm p \cdot \bm k (s - s') / m_X \right]
\approx 1 - i a \bm p \cdot \bm k (s - s') / m_X $.
Since 
$\int \bm k \cdot \nabla_{\bm p} f_X (\bm p) \, d^3 \bm p = 0$
due to spherical symmetry, only the second imaginary term
survives. Additionally, 
with the small relative velocity $|\bm v_{X c}| < 10^{-3}$, 
we can expand the second phase factor of \geqn{eq:delta_X_nf_fx}
also to the linear order,
$e^{- i a \bm v_{Xc} \cdot \bm k (s - s')} \approx
1 - i a \bm v_{X c} \cdot \bm k (s -s ')$. Their
product contains four terms. Besides the unit term,
only the product of the two imaginary
linear terms would give a real contribution,
$- a^2 (\bm p \cdot \bm k) (\bm v_{Xc} \cdot \bm k) (s - s')^2 / m_X$,
that can contribute an imaginary term to $\tilde \delta_X$,
\begin{align}
\hspace{-2mm}
  {\rm Im} \, \tilde \delta_X
=
  - 4 \pi G a
  \int_{s_i}^s ds'
  a^4(s') \rho_m \tilde \delta_m
  (\bm v_{X c} \cdot \bm k)
  (s - s')^2.
\label{eq:deltanu_clustering_cgf}
\end{align} 
With integration by part, one may prove that
$\int \frac{d^3 \bm p }{(2 \pi)^3} \frac{\bm p \cdot \bm k}{m_X} 
\bm k \cdot \nabla_{\bm p} f_X (\bm p) = - |\bm k|^2 \rho_X / m_X^2$.
The resulting $|\bm k|^2$ can be used to replace
the gravitational potential,
$|\bm k|^2 \tilde \Psi (s') = - 4 \pi G a^2 (s') \rho_m (s') \tilde \delta_m(s')$,
with the total matter density $\rho_m$ and overdensity $\tilde \delta_m$.

Below we will mainly use the bright galaxy sample (BGS)
category at redshift $z < 0.5$
where the relative velocity does not evolve with time
\cite{Zhu:2013tma,Inman:2016prk}.
So we can take the velocity $\bm v_{Xc}$ outside the integral. 
In addition, the matter overdensity $\tilde \delta_m$
follows the linear growth rate $D_+ = a$, and the matter
density $\rho_m \propto a^{-3}$. Consequently,
it is equivalent to replace the term
$a^2(s') \rho_m(s') \tilde \delta_m(s')$
inside the integration of \geqn{eq:deltanu_clustering_cgf} by
$a^2(s) \rho_m(s) \tilde \delta_m(s)$ that
can be removed from the integration.
The imaginary part of the total matter density $\tilde \delta_m$
can be parameterized as a phase $\tilde \phi_X$,
\begin{align}
  \tilde \phi_X
=
  - 4 \pi G \rho_{X 0}
  (\bm v_{X c} \cdot \bm k )
  \int_{s_i}^s ds' a^2(s') (s - s')^2,
\label{eq:def_fkf_mian}
\end{align}
where $\rho_{X 0} = \rho_X a^3$ is the current light DM
energy density.
In the final step, we have implemented the relation
${\rm Im}\,\tilde \delta_X \approx (\rho_X / \rho_m) {\rm Im}\,\tilde \delta_m$
for $\rho_X \ll \rho_m$.

Comparing with \geqn{eq:phichi1_main} that is obtained
in the free-streaming limit, the clustering limit
in \geqn{eq:def_fkf_mian} has quite
different features. Especially, the CGF has no explicit
mass dependence but implicitly
included in the relative velocity
$\bm v_{X c} \cdot \bm k \propto 1 / m^2_X$.
The projected sensitivity of CGF on
the light DM fraction $F_X$ will deteriorate for a
much larger mass, $m_X \gg \mathcal O(1)$\,eV,
with $1/m^2_X$ dependence.

In between, we can bridge
\geqn{eq:phichi1_main} and \geqn{eq:def_fkf_mian},
\begin{align} 
  \tilde \phi_X 
\equiv
- 4 \pi G F_X \rho_{\rm DM 0}
  (\bm v_{X c} \cdot \bm k )
  % k v_k
  g(|\bm k|),
\label{eq:def_int_A}
\end{align} 
with an interpolation function $g(|\bm k|)$, in the
similar ways as the one for the real part \cite{Chen:2020kxi}.
For convenience, we parameterize the energy density
$F_X \equiv \rho_{X 0} / \rho_{\rm DM 0}$
as fraction of the current total DM energy density
$\rho_{\rm DM 0}$ contained in dark matter $X$.
One choice of $g(|\bm k|)$ is,
\begin{align} 
  g(|\bm k|)
\equiv
  A
  \frac{k_{\rm fs}^3 }{ \left( |\bm k| + k_{\rm fs} \right)^3 }
+
  (B - A)
  \frac{k_{\rm fs}^4 }{ ( |\bm k|^2 + k_{\rm fs})^4 },
\label{eq:gk_interp_AB}
\end{align} 
such that $g(|\bm k|) = A \, k^3_{\rm fs} / |\bm k|^3$ or $B$
and \geqn{eq:def_int_A} reduces to the free-streaming
($|\bm k|^{-1} \ll k_{\rm fs}^{-1}$) or the clustering
($|\bm k|^{-1} \gg k_{\rm fs}^{-1}$) limit solution
in \geqn{eq:phichi1_main} or \geqn{eq:def_fkf_mian},
respectively. The corresponding coefficient $A$ ($B$)
is given by
\begin{subequations}
\begin{align} 
  A 
& \equiv
  \frac{\sqrt \pi}{ 3 }
  \frac{a^2}{k_{\rm fs}^3}
  \left( \frac{m_X}{T_{A0}} \right)^3
  \frac{e^{-y_i}}{ \sqrt{y_i} },
\label{eq:def_A_phiX}
\\
  B
& \equiv
  \int_{s_i}^s ds'
  a^2(s') (s - s')^2.
\label{eq:def_AB_phiX}
\end{align}
\end{subequations}

\vspace{2mm}
{\bf Galaxy Cross Correlation and Projected Sensitivity} --
Neither the light DM $X$ nor the major CDM component can be
directly observed. Fortunately, galaxies formation is
influenced by the gravitational potential of the total matter
including both $X$ and CDM. It is possible to use
the galaxy distribution to reconstruct the matter density
distribution. More specifically, the characteristic dipole
density induced by the CGF effect can be
reconstructed from the galaxy cross correlation
\cite{Zhu:2013tma,Ge:2023nnh}.

The galaxy number overdensity
$\tilde \delta_{g \alpha} 
= b_\alpha \tilde \delta_m 
+ i b_X \tilde \phi_X \tilde \delta_{\rm m} $
is a linear combination of the CDM ($\tilde \delta_m$)
and the light DM ($\tilde \phi_X \tilde \delta_m$)
overdensities. The type-$\alpha$ galaxy bias
$b_\alpha$ is for the CDM
and $b_X$ is for the light DM $X$.
Following the usual treatment, we take the same $b_X = 1$
as cosmic neutrinos \cite{LoVerde:2014pxa}. 

We define the observable signal as the imaginary
part of the galaxy cross correlation 
$ \mathcal S 
\equiv 
  {\rm Im} 
  \langle \tilde \delta_{g \alpha} \tilde \delta_{g \beta} \rangle  $ 
and, and noise as its variance
$\mathcal N \equiv \sqrt{ \langle \mathcal S \rangle^2 - \langle S \rangle^2   }$ \cite{Ge:2023nnh}.
The signal-to-noise ratio (SNR) is then given by, 
\begin{align}
  \left( \frac{\mathcal S}{\mathcal N} \right)^2
& = 
  \sum_{z_i} 
  \frac{\Delta b^2 V_i}{5 \pi^2}
  \int d|{\bm k}|
  \frac{|{\bm k}|^2 P_{m}^2}{\rm Det[C]} 
\left[ 
  \frac{\left\langle  \dot{\tilde \phi}_X^2\right\rangle} {H^2}
+
\right.
\label{eq:lnl_dpdppp}
\\
& 
\left.
    \left(f^2 + \frac{10} 3 f + 5 \right)  
    \langle \tilde \phi_X^2 \rangle 
  +2\left( f + \frac{5} 3 \right)
  \frac{\langle \tilde \phi_X \dot{\tilde \phi}_X  \rangle }{H}
\right],
\notag
\end{align}
where $\Delta b \equiv b_\alpha - b_\beta$,
$V_i$, $P_m$, ${\rm Det} [C]$, $\dot{\tilde \phi}_X$, 
and $f$ are the bias differences between two galaxy types, 
the effective survey volume, the matter power spectrum,
the determinant of the covariance matrix 
$C_{\alpha \beta} \equiv \langle \delta_{g \alpha} \delta_{g \beta} \rangle$,
the time derivative of 
field $\tilde \phi_X$ in \geqn{eq:phichi1_main}, and the growth rate 
$f \equiv d \ln D_+ / d \ln a \approx \Omega_m(z)^{0.55}$
where $\Omega_m(z)$ is the time-dependent matter fraction.
The ensemble averages $\langle \dot{\tilde \phi}_X^2 \rangle$, 
$\langle \tilde \phi_X^2 \rangle $, and 
$ \langle \tilde \phi_X \dot{\tilde \phi}_X \rangle$
can be directly derived from \geqn{eq:def_int_A}-\geqn{eq:def_AB_phiX},
and are functions of the light DM fraction $F_X$,
its mass $m_X$, and the spectrum parameter $T_A$,
with more details in the supplementary materials.
In the following, we implement our calculation by
using the CLASS code \cite{2011JCAP...07..034B,Lesgourgues_2011}.

\begin{figure}[t!]
\centering
\includegraphics[width=0.49\textwidth,height=57mm]{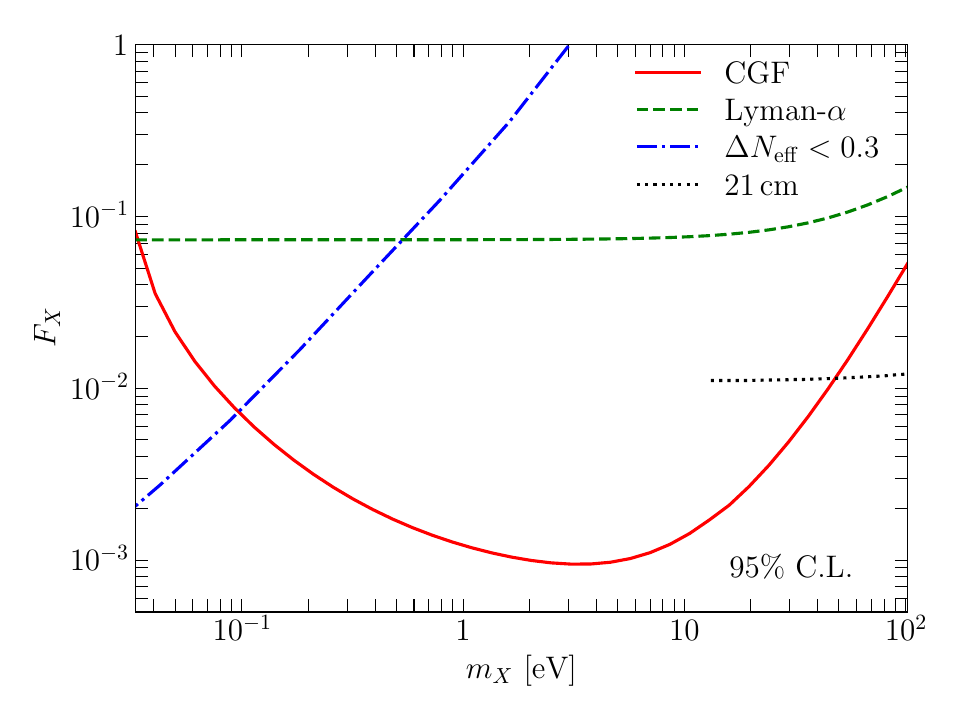}
\caption{
The projected CGF sensitivity (red solid) on the light DM
energy fraction $F_X \equiv \Omega_X / \Omega_{\rm DM}$
as function of the light DM mass $m_X$ from the
DESI observations of the BGS galaxy category.
For illustration, we take the phase space distribution
in \protect\geqn{eq:fp_freeze_in} of a typical
freeze-in DM with the current spectrum
parameter $T_{A 0} = 10^{-4}$\,eV around the Universe
temperature today. 
For comparison, the existing Lyman-$\alpha$ (green dashed)
CMB $\Delta N_{\rm eff}$ (blue dash-dotted) and
21\,cm (black dotted) constraints are also shown together.} 
\label{fig:fchi_mchi}
\end{figure}

Using the DESI catalogs \cite{Ge:2023nnh} 
for the BGS and faint galaxies \cite{DESI:2016fyo},
the projected sensitivity on the light DM fraction
$F_X$ as function of its mass $m_X$
is plotted as the red solid line in \gfig{fig:fchi_mchi}. 
For $m_X < 1\,{\rm eV}$ where the observable scale
is below the free-streaming scale,
the solution \geqn{eq:phichi1_main} works well.
As already analyzed above, the projected sensitivity
becomes stronger for a heavier mass with $m_X^2$
scaling behavior. On the other hand, for $m_X > 10$\,eV
where the observable scale is much larger than its 
free-streaming scale $k_{\rm fs}^{-1} \lesssim 0.384\,{\rm Mpc}/ h$,
the solution follows \geqn{eq:def_int_A} with a coefficient
$g(|\bm k|) = B$ in \geqn{eq:def_AB_phiX}. 
With $1/m^2_X$ mass dependence, 
the light DM fraction $F_X$ receives weaker constraint
for heavier mass. So the intermediate region
$1\,{\rm eV} < m_X < 10\,{\rm eV}$
exhibits the strongest constraint, $ F_X < 10^{-3}$.
The overall features of the projected sensitivity
can be readily understood from the mass scaling behaviors.

Our projected sensitivity from CGF in \gfig{fig:fchi_mchi}
comes from a conservative estimation without considering
the cosmic neutrinos. Especially, for the left part
$m_X \lesssim 0.1$\,eV where the neutrino mass sum starts
to become comparable with $m_X$, the cosmic neutrinos 
can also contribute to the dipole distribution
\cite{Ge:2023nnh,Ge:2024kac}.
By substracting the the cosmic neutrino contribution,
the constraints on the light DM $X$ is expected to
become stronger than the red solid line in \gfig{fig:fchi_mchi}.
For larger $m_X \gg 0.1$\,eV, the neutrino contribution
can be safely ignored.

For comparison, we also show the Lyman-$\alpha$ (green
dashed) and $\Delta N_{\rm eff}$ (blue dash-dotted) constraints
in \gfig{fig:fchi_mchi}. Since the light DM $X$
has a higher velocity than the major CDM, it
suppresses the structure formation below its free-streaming scale.
Such effect can be probed by the Lyman-$\alpha$ observations
that is sensitive to scales 
$ 0.5\,h/{\rm Mpc} < |\bm k| < 20\,h/{\rm Mpc}$ \cite{Murgia:2017lwo}.
This suppression effect can be parameterized by
a transfer function $\mathcal T (|\bm k|)$,
$ \tilde \delta_m = \mathcal T (|\bm k|) \tilde \delta_c $,
where $\tilde \delta_c$ is the CDM overdensity.
Using a fitting function for $\mathcal T(|\bm k|)$
from N-body simulations \cite{Kamada:2016vsc}
and adopting a conservative constraint 
$ \mathcal T^2 ( |\bm k| < 20\,h/{\rm Mpc}) \geq 0.7 $ \cite{Carena:2021bqm}
taken from Fig.8 of \cite{Murgia:2017lwo},
we plot the resulting Lyman-$\alpha$ constraint as
the green dashed line in \gfig{fig:fchi_mchi}. Across almost
the whole mass region in \gfig{fig:fchi_mchi},
the Lyman-$\alpha$ constraint is almost flat
$F_X < \mathcal O (0.1)$ \cite{Hooper:2022byl}.
This is because the light
DM suppresses the power spectrum with a
fraction $8 F_X$ relative the original
power spectrum \cite{Hu:1997mj},
which is blind to the light DM mass $m_X$.

For mass below 0.1\,eV, the light DM $X$ remains
relativistic during the recombination.
This extra radiation energy density can affect the
CMB through the ISW (Integrated Sachs–Wolfe) effect
\cite{Brust:2013ova}, which is usually parameterized
as $\Delta N_{\rm eff}$ in unit of the
effective number of neutrino species. 
We plot the constraint for $\Delta N_{\rm eff} \lesssim 0.3$
\cite{ParticleDataGroup:2022pth,Planck:2018vyg},
as the blue dash-dotted line in \gfig{fig:fchi_mchi}.
This CMB sensitivity decreases with the light DM mass
$m_X$ very quickly.

The CGF constraint
is stronger than the existing Lyman-$\alpha$
and $\Delta N_{\rm eff}$ constraints for
$ 0.1\,{\rm eV} < m_X < 100\,{\rm eV}$.
Notably, the CGF sensitivity can be two orders stronger
in the middle region $ 1\,{\rm eV} < m_X < 10\,{\rm eV}$.
Even the forecasted sensitivity for the
future 21-cm observation can reach only $F_X \lesssim 10^{-2}$
(shown as the black dotted line in \gfig{fig:fchi_mchi})
\cite{Giri:2022nxq}.

\vspace{2mm}
{\bf Conclusion and Discussions} -- 
Although the large scale structure of our Universe
prefers the cold DM, a light species 
can still exist as minor component so long as
its energy fraction is small enough. However, this
leads to an imaginable difficulty of probing
such minor light DM component.

Fortunately, if the minor light DM and the major
cold DM have quite different masses, they would develop
relative bulk velocity. Consequently, the light DM fluid
would flow by the cold DM halos and the gravitational
attraction between them would lead to the cosmic
gravitational effect in the same way as the cosmic
neutrino fluid. However, the light DM with larger
mass would have much stronger effect than its neutrino
counterparts. This makes the CGF an ideal
tool for probing the light DM.

We provide analytical understanding of
the sensitivity scaling behaviors with the
light DM mass $m_X$ in both the free-streaming
($m_X \lesssim 1$\,eV) and clustering
($m_X \gtrsim 10$\,eV) limits.
Our study shows that for the light DM mass
$1\,{\rm eV} < m_X < 10$\,eV, the projected
CGF sensitivity with the DESI observation
can reach $F_X \lesssim 10^{-3}$ which
stronger than the existing Lyman-$\alpha$
and CMB $\Delta N_{\rm eff}$ constraints by
two orders. With the upcoming galaxy surveys, 
such as the spectroscopic survey DESI
\cite{DESI:2016fyo,DESI:2018ymu}
as well as the photometric surveys like
LSST \cite{2009arXiv0912.0201L}, 
WFIRST \cite{WFIRST:2018mpe}, Euclid \cite{EUCLID:2011zbd}, 
and CSST \cite{cao2018testing,Cao:2021bqm,Cao:2021ykj}, 
we expect the CGF effect to receive emerging
firm data.

\section*{Acknowledgements}

The authors are supported by the National Natural Science
Foundation of China (12425506, 12375101, 12090060 and 12090064) and the SJTU Double First
Class start-up fund (WF220442604).
SFG is also an affiliate member of Kavli IPMU, University of Tokyo.

\bibliographystyle{utphys}
\bibliography{dmCosmicFocusing}

\onecolumngrid
\appendix

\clearpage

\setcounter{equation}{0}
\setcounter{figure}{0}
\setcounter{table}{0}
\setcounter{page}{1}
\makeatletter
\renewcommand{\theequation}{S\arabic{equation}}
\renewcommand{\thefigure}{S\arabic{figure}}
\renewcommand{\thepage}{S\arabic{page}}

\begin{center}
\textbf{\large Supplemental Material for the Letter\\[0.5ex]
{\em Probing Light Dark Matter with Cosmic Gravitational Focusing}}
\end{center}

In this Supplemental Material, we provide explicit derivations 
for the CGF signal, especially those ensemble averages.
Since the variations for constructing the signal-noise-ratio in
\geqn{eq:lnl_dpdppp} contains both ${\tilde \phi}_X$ and
its time derivative $\dot{\tilde \phi}_X$ which in turn
can be expressed as functions of the relative velocity
variance (dispersion) as well as the interpolation coefficients
$A$ and $B$, we will first explore the velocity dispersions
in Sec.\,A and then the interpolation coefficients in
Sec.\,B. More tedious derivations can be found in Sec.\,C.

\section{A.~ Relative Velocity}
\label{sec:vDispersion}

The average relative velocity $\bm v_{Xc}$
between the light DM $X$ and
the major CDM component can be estimated as its dispersion 
(similar as the cosmic neutrino case \cite{Zhu:2013tma,Ge:2023nnh}),
\begin{align} 
  \langle \bm v_{X c}^2 \rangle  
\equiv 
  \int \frac{d |\bm k|}{|\bm k|} 
  \Delta_\zeta^2 (\bm k)
  \frac{ | T_{\theta_{X c}}|^2 }{|\bm k|^2}
  \left| \widetilde W (|\bm k| R) \right|^2,
\label{eq:v_nuc2}
\end{align} 
where $\Delta_\zeta$, $\widetilde W(|\bm k| R)$, $T_{\theta_{X c}}$
are the dimensionless primordial power spectrum,
the window function filter with scale $R$, 
and the transfer function of the relative velocity,
respectively. We choose the filter scale at
$R = 5\,{\rm Mpc/h}$ \cite{Ge:2023nnh} in our calculation. 
The transfer function
$ T_{\theta_{X c}} \equiv (\tilde\theta_X - \tilde\theta_c)/\zeta (\bm k) $
can be directly obtained from the 
\href{https://lesgourg.github.io/class_public/class.html}{CLASS}
code \cite{2011JCAP...07..034B,Lesgourgues_2011} simulation.

\begin{figure}[h]
\centering
\includegraphics[width=0.49\textwidth]{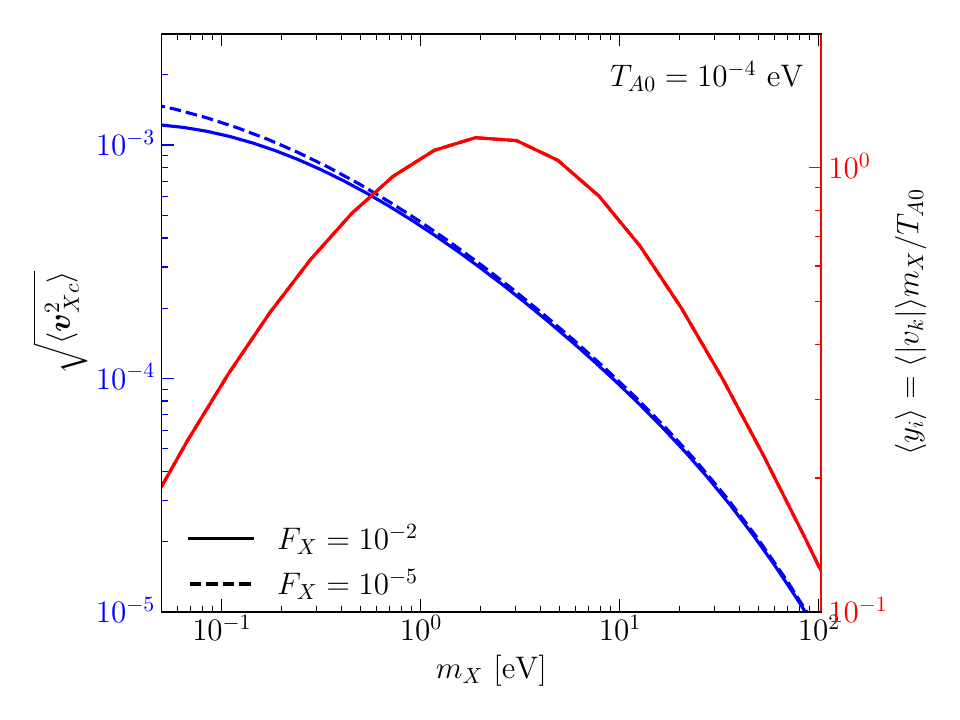}
\includegraphics[width=0.49\textwidth]{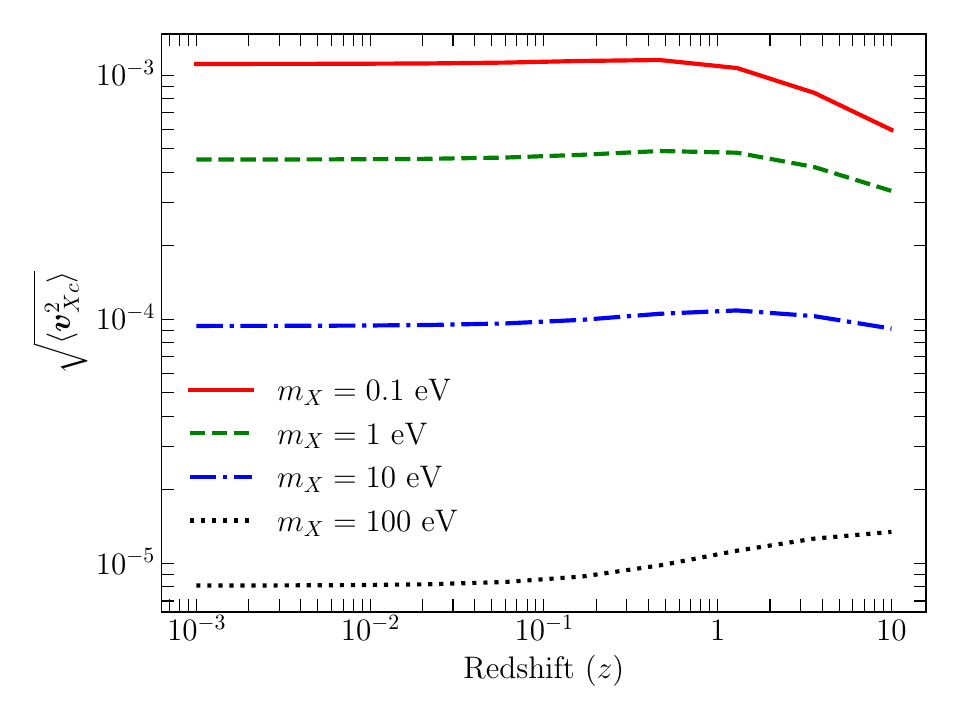}
\caption{
{\bf Left:}
The velocity variance $\sqrt{\langle \bm v_{X c}^2 \rangle}$ (blue and the left vertical axis)
and the expansion parameter $\langle y_i \rangle $ (red and the right vertical axis) varying
with the light DM mass $m_X$ for $F_X = 10^{-2}$ 
(solid) and $F_X = 10^{-5}$ (dashed) given a typical spectrum parameter
today $T_{A 0} = 10^{-4}\,{\rm eV}$.
{\bf Right:} Relative velocity evolution of
the freeze-in DM $X$ for
$m_X = 0.1$\,eV, 1\,eV, 10\,eV, and 100\,eV.
}
\label{fig:relV_WDM_CDM}
\end{figure}

In the CLASS code \cite{2011JCAP...07..034B,Lesgourgues_2011},  
we specify the light DM mass $m_X$ and energy fraction
$\Omega_X \equiv \rho_{X 0} / \rho_c$,
where $\rho_{X 0}$ is the light DM density today
and $\rho_c$ is the critical density,
while fixing the total DM energy fraction
$\Omega_{\rm DM} h^2 = (\Omega_{\rm CDM} + \Omega_X) h^2 = 0.12$  
according to the Planck 2018 data \cite{Planck:2018vyg}.
By varying the light DM mass $m_X$ and its fraction
$F_X \equiv \Omega_X / \Omega_{\rm DM}$ relative to the
total DM energy fraction $\Omega_{\rm DM}$,
we illustrate the velocity dispersion
$\sqrt{ \langle \bm v_{Xc}^2 \rangle}$ and its time evolution
in the left and right panels of \gfig{fig:relV_WDM_CDM},
respectively. Note that the freeze-in phase space distribution
is taken from \geqn{eq:fp_freeze_in}. While the left panel
shows that the velocity dispersion
$\sqrt{ \langle \bm v_{Xc}^2 \rangle}$ decreases with the
light DM mass $m_X$, the right panel
demonstrates that the relative velocity remains nearly
constant with time for $z<1$.

This velocity evaluated in \geqn{eq:v_nuc2} is
specifically applicable to small scales. However,
as scale increases, the relative velocity is expected 
to decrease because the velocity field is not coherent on large
scales, which is verified by N-body simulations \cite{Inman:2016prk}.
To account for this decreasing behavior,
we incorporate the $\Theta(|\bm k| - |\bm k'|)$
function \cite{Okoli:2016vmd,Ge:2023nnh} within \geqn{eq:v_nuc2}. 

\begin{figure}[!t]
\centering
\includegraphics[width=0.62\textwidth]{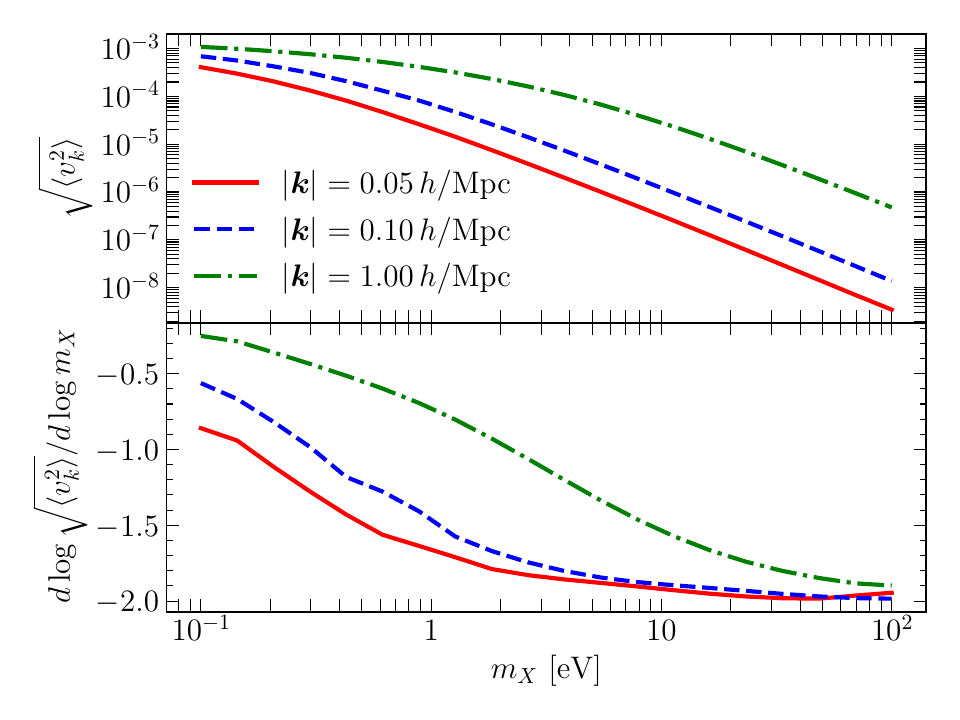}
\caption{
{\bf Upper:} The mass scaling behavior of the velocity dispersion,
$\sqrt{\langle v_k^2 \rangle } \propto m_X^n$
as power of the light DM mass $m_X$
and {\bf Lower:} the power index
$n \equiv d \left( \log \sqrt{ \langle v_k^2 \rangle  } \right) / d \log m_X $.
}
\label{fig:vk2_mXn}
\end{figure}

As we will show below, the velocity dependence
of the ensemble averages of \geqn{eq:lnl_dpdppp} all
appears in terms of $v_k \equiv \bm v_{Xc} \cdot \hat{\bm k}$
where $\hat{\bm k}$ is the unit vector of the wave number $\bm k$.
In particular, there are just three independent forms,
\begin{subequations}
\begin{align} 
  \langle v_k^2 \rangle
= &
  \frac 1 3
  \int \frac{d |\bm k'|}{|\bm k'|}
  \Theta (|\bm k| - |\bm k'|) 
  \left| \widetilde W (|\bm k'| R) \right|^2
  \Delta_{\zeta}^2
  \left| \frac{T_{\theta_{X c}} (\bm k', z)}{|\bm k'|} \right|^2,
\label{eq:vnuc_Theta}
\\
  \langle v_k \partial_z v_k \rangle
= & 
  \frac 1 3
  \int \frac{d |\bm k'|}{|\bm k'|}
  \Theta (|\bm k| - |\bm k'|)  
  \left| \widetilde W (|\bm k'| R) \right|^2
  \Delta_{\zeta}^2 (\bm k')
\left[ 
  \frac{T_{\theta_X c}}{|\bm k'|}
  \frac{\partial_z T_{\theta_X c}}{|\bm k'|}
\right],
\\
  \langle \partial_z v_k \partial_z v_k \rangle
= &
  \frac 1 3
  \int \frac{d |\bm k'|}{|\bm k'|}
  \Theta (|\bm k| - |\bm k'|)
  \left| \widetilde W (|\bm k'| R) \right|^2
  \Delta_{\zeta}^2 (\bm k')
\left[ 
  \frac{\partial_z T_{\theta_\nu c}}{|\bm k'|}
\right]^2.
\end{align}
\label{eq:coherent_velocity}
\end{subequations}
In addition to velocity dispersion
$\sqrt{\langle v^2_k \rangle}$,
we have also plotted the power index, 
$n \equiv d \left( \log \sqrt{ \langle v_k^2 \rangle  }  \right) / d \log m_X $
in \gfig{fig:vk2_mXn}.
For $m_X = 0.1\,{\rm eV}$ and $100\,{\rm eV}$, 
the value of $n$ can be extracted from the curve
slope to be $n \approx -0.5$ and $-2$, respectively,
over the $|\bm k|^{-1}$ range of $(1, 20)\,{\rm Mpc}/h$.
Equivalently, $\sqrt{\langle v_k^2 \rangle}$ scales with
the light DM mass $m_X$ roughly as
$\sqrt{ \langle v_k^2 \rangle} \propto m^n_X$.

\section{B.~ Interpolation Coefficients between Free-Streaming and Clustering Limits}

The coefficient $A$ in \geqn{eq:def_A_phiX} for
the light DM $X$ can be written as,
\begin{align} 
  A
\approx
  \frac{\sqrt \pi} 3
  \frac{1}{k_{\rm fs}^3}
  \left( \frac{m_X}{T_{A 0}} \right)^{5/2} 
  a^{3/2}
  \frac{ 1 }{ \sqrt{ | v_k | } },
\label{eq:phichi_vk_non}
\end{align} 
where we have replaced
$y_i \equiv m_X |\bm v_{Xc} \cdot \hat{\bm k}| / T_A$
to give that $a^{3/2}$ factor.
In most of the mass region, the parameter 
$ \langle y_i \rangle 
=  m_X \langle | v_k |  \rangle / T_A < 1$ 
by choosing the smallest scale $|\bm k|^{-1} = 1\,{\rm Mpc}/h$
in our calculation,
as shown in the left panel of \gfig{fig:relV_WDM_CDM}.
So we can approximate the exponential factor as
$e^{- y_i} \approx 1$. In the middle region,
where $ y_i > 1 $, we can just add the factor
$e^{ - \langle y_i \rangle}$ back into the expression
for the coefficient $A$.

The time derivative of $v_k A$ can be derived as,
\begin{align}
  \frac{d (v_k A) }{ dt}
& =
  \frac{\sqrt \pi} 3
  \frac{1}{k_{\rm fs}^3}
  \left( \frac{m_X}{T_{A 0}} \right)^{5/2} 
  \frac{d}{dt}
  \left( a^{3/2} \frac{ v_k }{ \sqrt{ | v_k | } }   \right),
\end{align}
Then, using $ d a^{3/2} / dt = \frac 3 2 a^{1/2} \dot a
= \frac 3 2 a^{3/2} H$,
where $H \equiv \dot a / a$ is the Hubble rate,
the above equation becomes,
\begin{align}
  \frac{d (v_k A) }{ dt}
& =
  \frac{\sqrt \pi} 3
  \frac{1}{k_{\rm fs}^3}
  \left( \frac{m_X}{T_{A 0}} \right)^{5/2} 
  \left( 
    \frac{3 a^{3/2}} 2  H 
    \frac{ v_k }{ \sqrt{ | v_k | } }
  +  a^{3/2} 
    \frac 1 { 2 \sqrt{|v_k|} }
    \frac{d v_k}{ d t }
  \right),
\end{align}
where we have used the fact that
$\partial_t (v_k / \sqrt{|v_k|})$
can be replaced by
$\partial_{v_k} (v_k / \sqrt{|v_k|}) \partial_t v_k$
which gives $\dot v_k / 2 \sqrt{|v_k|}$
with $\dot v_k \equiv d v_k /dt$.
During the derivation, we have used
$\partial_{v_k} |v_k| = |v_k| / v_k$ for $v_k \neq 0$.
 
We can change the time $t$ to redshift $z$
by using $1+z = 1/a$ and $d/dt = - (1+z) H d/dz$, 
\begin{align}
  \frac{d (v_k A) }{ dt}
& =
  \frac {\sqrt \pi} 3
  \frac{1}{k_{\rm fs}^3}
  \left( \frac{m_X}{T_{A 0}} \right)^{5/2} 
  H
  \left[ 
    \frac{3 }{2 (1+z)^{3/2} }
    \frac{ v_k }{ \sqrt{ | v_k | } }
  -
    \frac{1 }{2 \sqrt{1+z}}
    \frac{ 1  }{ \sqrt{ |v_k| }} 
    \frac{d v_k}{d z}
  \right].
\label{eq:dvkA_dt}
\end{align}

From \geqn{eq:def_AB_phiX}, we can get,
\begin{align} 
  \dot B 
=
  \frac{1}{a^2}
  \frac{d B}{d s}
=  
  \frac 2 {a^2}
  \int_{s_i}^s ds'
  a^2(s')  
  (s - s'),
\quad \mbox{and} \quad
  \partial_z B 
=  
- \frac 2 {a H}
  \int_{s_i}^s ds' a^2(s')   (s - s').
\label{eq:B}
\end{align} 
Then the redshift derivative of $v_k B$ can be
expressed as,
\begin{align} 
  \frac{d (v_k B)}{ d t } 
=
  - (1+z) H
  \frac{d (v_k B)}{ d z } 
=
  - (1+z) H
  \left(  \partial_z v_k B + v_k \partial_z B  \right).
\label{eq:dvkb_dt}
\end{align}

\section{C.~ Variance}

As elaborated in the main text, the phase $\tilde \phi_X$
is obtained from the interpolation between the free-streaming
and clustering limits. For convenience, we show the complete
form of \geqn{eq:lnl_dpdppp} by combining \geqn{eq:def_int_A} and \geqn{eq:gk_interp_AB},
\begin{align} 
  \tilde \phi_X 
\equiv
  - 4 \pi G F_X \rho_{\rm DM 0}
  (\bm v_{X c} \cdot \bm k)
  \left[ 
    A
    \frac {k_{\rm fs}^3}
          { \left( |\bm k| +  k_{\rm fs} \right)^3 }
  +
    (B - A)
    \frac {k_{\rm fs}^4 }
          { ( |\bm k| + k_{\rm fs})^4 }
  \right]. 
\label{eq:tilde_phiX}
\end{align} 
Below we will try to show the explicit forms of
$\langle \tilde \phi_X^2 \rangle $,
$\langle \dot{\tilde \phi}_X^2 \rangle $,
and $\langle \tilde \phi_X  \dot{\tilde \phi}_X \rangle $,
respectively

\vspace{3mm}
\noindent
{\bf (C-I): Variance of $\langle \tilde \phi_X^2 \rangle $:}
Using \geqn{eq:tilde_phiX}, the ensemble average
of $\tilde \phi_X^2$ can be expressed as,
\begin{align} 
  \langle \tilde \phi_X^2 \rangle  
=
 \left( 4 \pi G F_X \rho_{\rm DM 0} \right)^2
 |\bm k|^2
  \left[ 
    \langle v_k^2 A^2 \rangle
    \frac{  k_{\rm fs}^6 }{ \left( |\bm k| +  k_{\rm fs} \right)^6 }
  + 2  \langle v_k^2 A ( B - A ) \rangle
    \frac{  k_{\rm fs}^7 }{ \left( |\bm k| +  k_{\rm fs} \right)^7 }
  +
    \langle v_k^2 (B - A)^2 \rangle
    \frac{  k_{\rm fs}^8 }{ ( |\bm k| + k_{\rm fs})^8 }
  \right],
\label{eq:phi_non_x2}
\end{align} 
where we have used the simplified notation
$v_k \equiv \bm v_{X c} \cdot \hat{\bm k}$.
Below we will calculate the three ensemble averages
in \geqn{eq:phi_non_x2} one by one.

\vspace{2mm}
{\bf (1):} The first ensemble average term in
\geqn{eq:phi_non_x2} can be written as,
\begin{align} 
  \langle v_k^2 A^2 \rangle  
=
  \frac{\pi}{ 9 }
  \frac{1}{k_{\rm fs}^6}
  \left( \frac{m_X}{T_{A 0}} \right)^{5} 
  a^{3}
  \left\langle 
    \frac{ v_k^2 }{ | v_k | }
  \right\rangle
=
  \frac{\sqrt{2 \pi}}{ 9 }
  \frac{1}{k_{\rm fs}^6}
  \left( \frac{m_X}{T_{A 0}} \right)^{5} 
  a^{3}
  \sqrt{ \langle v_k^2 \rangle  },
\label{eq:avg_vkA2}
\end{align} 
with the explicit form of $A$ in \geqn{eq:phichi_vk_non}.
Being a Gaussian random distribution $x$, the
relative velocity esemble average can be simplified,
\begin{align} 
  \left\langle \frac{x^2}{|x|} \right\rangle  
=
  \langle |x| \rangle 
=
  \sqrt{\frac{2}{\pi}}
  \sqrt{\langle x^2 \rangle}.
\label{eq:avg_x2_absx}
\end{align}

\vspace{2mm}
{\bf (2):} The second ensemble average term in \geqn{eq:phi_non_x2} can be written as,
\begin{align} 
  \langle v_k^2 A ( B - A ) \rangle
=
  \langle v_k^2 A\rangle  B  
-
  \langle v_k^2 A^2 \rangle,
\end{align}
where the $\langle v_k^2 A^2 \rangle $ has already
been derived in \geqn{eq:avg_vkA2}. Since the $B$ in
\geqn{eq:B} does not have dependence on the relative
velocity $v_k$, it can be directly moved outside of
the esemble average. Then we only need to expand
$\langle v^2_k A \rangle$.

Putting the coefficient $A$ of \geqn{eq:phichi_vk_non}
back into the first term,
\begin{align} 
  \langle v_k^2 A \rangle
=
  \frac{\sqrt \pi}{ 3 }
  \frac{1}{k_{\rm fs}^3}
  \left( \frac{m_X}{T_{A 0}} \right)^{5/2} 
  a^{3/2}
  \left\langle 
    \frac{ v_k^2 }{ \sqrt{ | v_k | } }
  \right\rangle
=
  \frac{\sqrt \pi}{ 3 }
  \frac{1}{k_{\rm fs}^3}
  \left( \frac{m_X}{T_{A 0}} \right)^{5/2} 
  a^{3/2}
  \frac{ 2^{3/4}  \Gamma(5/4) }{ \sqrt{\pi} }
  \langle v_k^2 \rangle^{3/4}.
\label{eq:avg_vkA}
\end{align} 
Here, we also used the property of a Gaussian
variable $x \equiv v_k$,
\begin{align} 
  \left\langle \frac{x^2}{ \sqrt{|x|} } \right\rangle  
=
  \langle |x|^{3/2}  \rangle 
=
  \frac{ 2^{3/4}  \Gamma(5/4) }{ \sqrt{\pi} }
  \langle x^2 \rangle^{3/4}.
\label{eq:avg_x2_sqx}
\end{align}

{\bf (3):} The third ensemble average term in \geqn{eq:phi_non_x2} can be written as,
\begin{align} 
  \langle v_k^2 (B - A)^2 \rangle
=
  \langle v_k^2 \rangle B^2
-
  2 \langle v_k^2 A \rangle B
+
  \langle v_k^2  A^2 \rangle
\end{align} 
where the ensemble averages involving $A$
have already been derived in
\geqn{eq:avg_vkA2} and \geqn{eq:avg_vkA}
while $\langle v^2_k \rangle$ can be
found in \geqn{eq:vnuc_Theta}.

\vspace{3mm}
\noindent
{\bf (C-II): Variance of $\langle \dot{\tilde \phi}_X^2 \rangle $:}
We need to first derive the time derivative of the
phase $\tilde \phi_X$ in \geqn{eq:tilde_phiX},
\begin{align} 
  \dot{\tilde \phi}_X 
=
  - 4 \pi G F_X \rho_{\rm DM 0}
  |\bm k|
  \left[ 
    \frac{d (v_k A ) }{d t}
    \frac{  k_{\rm fs}^3 }{ \left( |\bm k| +  k_{\rm fs} \right)^3 }
  +
    \frac{ d [ v_k (B - A) ] }{d t}
    \frac{  k_{\rm fs}^4 }{ ( |\bm k| + k_{\rm fs})^4 }
  \right]. 
\label{eq:tilde_dphiX}
\end{align} 
Then, the variation of $\dot{\tilde \phi}_X$ becomes,
\begin{align} 
  \langle \dot{\tilde \phi}_X^2 \rangle  
& =
  \left( 4 \pi G F_X \rho_{\rm DM 0} \right)^2
  |\bm k|^2
  \left[ 
    \left\langle
      \frac{d (v_k A ) }{d t}
      \frac{d (v_k A ) }{d t}
    \right\rangle
    \frac{  k_{\rm fs}^6 }{ \left( |\bm k| +  k_{\rm fs} \right)^6 }
   +
    2
    \left\langle
      \frac{d (v_k A ) }{d t}
      \frac{ d [ v_k (B - A) ] }{d t}
    \right\rangle
    \frac{  k_{\rm fs}^7 }{ \left( |\bm k| +  k_{\rm fs} \right)^7 }
  \right.
\nonumber
\\
&
\hspace{33mm}
\left.
  +
    \left\langle
      \frac{ d [ v_k (B - A) ] }{d t}
      \frac{ d [ v_k (B - A) ] }{d t}
    \right\rangle
    \frac{  k_{\rm fs}^8 }{ ( |\bm k| + k_{\rm fs})^8 }
  \right].
\label{eq:dphi_non_x2}
\end{align}

{\bf (1):} 
Using the result \geqn{eq:dvkA_dt}, 
the first ensemble average term in \geqn{eq:dphi_non_x2} can be written as
\begin{align} 
  \left\langle
    \frac{d (v_k A ) }{d t}
    \frac{d (v_k A ) }{d t}
  \right\rangle
& =
  \frac{\pi}{ 9 }
  \frac{1}{k_{\rm fs}^6}
  \left( \frac{m_X}{T_{A 0}} \right)^{5} 
  H^2
  \left[ 
    \frac{9 }{4 (1+z)^3 }
    \left\langle \frac{ v_k^2 }{ | v_k | } \right\rangle
  -
    \frac{ 3 }{2 (1+z)^2 }
    \left\langle \frac{ v_k \partial_z v_k  }{ |v_k| }  \right\rangle
  +
    \frac{1 }{4 (1+z) }
    \left\langle \frac{ ( \partial_z v_k )^2  }{ |v_k| }  \right\rangle
  \right].
\label{eq:dvkAdvkA_t}
\end{align} 
The first term in \geqn{eq:dvkAdvkA_t} can be obtained
from the properties of the Gaussian variable $x = v_k$
in \geqn{eq:avg_x2_absx}. For the second term,
the redshift derivative $\partial_z$ on the single
velocity $v_k$ can be moved to be an overall one,
\begin{align} 
  \partial_z \left(  \frac{v_k^2}{ |v_k| } \right)  
& =
  \frac{2 v_k \partial_z v_k}{|v_k|}
+ 
  v_k^2 \partial_z \frac 1 {|v_k|}
=
  \frac{2 v_k \partial_z v_k}{|v_k|}
- 
  v_k^2 
  \frac 1 {|v_k|^2}
  \frac{d |v_k|}{d v_k}
  \partial_z v_k 
=
  \frac{ v_k \partial_z v_k}{|v_k|}.
\end{align} 
Then, the second term in \geqn{eq:dvkAdvkA_t} can be written as,
\begin{align} 
  \left\langle  \frac{ v_k \partial_z v_k}{|v_k|} \right\rangle 
=
  \left\langle  \partial_z  \frac{v_k^2}{ |v_k| }   \right\rangle 
=
  \partial_z 
  \left\langle  \frac{v_k^2}{ |v_k| }   \right\rangle 
=
  \sqrt{ \frac 2 {\pi}}
  \partial_z
  \sqrt{ \langle  v_k^2  \rangle  } 
=
  \sqrt{ \frac 2 {\pi}}
  \frac{\partial_z \langle  v_k^2  \rangle}{ 2 \sqrt{ \langle v_k^2 \rangle  }  }
=
  \sqrt{ \frac 2 {\pi}}
  \frac{\langle  v_k \partial_z v_k \rangle}{  \sqrt{ \langle v_k^2 \rangle  }  }.
\label{eq:pvkabsvk_new}
\end{align} 
In the last equality, the order of $\partial_z$ and
the ensemble average $\langle \dots \rangle$ has
been exchanged, since the time derivative and
coordinate average is independent of each other,
i.e., 
$ \partial_z \langle v_k^2 \rangle  
= \langle \partial_z v_k^2  \rangle  
= 2 \langle v_k \partial_z v_k \rangle $.
Both the numerator and denominator of \geqn{eq:pvkabsvk_new}
have been shown in \geqn{eq:coherent_velocity}.
Similarly, the additional ensemble average
of the third term in \geqn{eq:dvkAdvkA_t} can 
be written as,
\begin{align} 
  \left\langle \frac{\partial_z v_k \partial_z v_k}{ |v_k| } \right\rangle  
=
  \sqrt{\frac 2 \pi}
  \frac{ \langle  \partial_z v_k \partial_z v_k  \rangle   }{ \sqrt{ \langle v_k^2 \rangle  } }.
\label{eq:llfcvkr}
\end{align}

Then, putting \geqn{eq:pvkabsvk_new} and \geqn{eq:llfcvkr}
back into the ensemble average \geqn{eq:dvkAdvkA_t},
\begin{align} 
  \left\langle
    \frac{d (v_k A ) }{d t}
    \frac{d (v_k A ) }{d t}
  \right\rangle
=
  \frac{\sqrt{2 \pi}}{ 9 }
  \frac{1}{k_{\rm fs}^6}
  \left( \frac{m_X}{T_{A 0}} \right)^{5} 
  H^2
  \left[
    \frac{9 }{4 (1+z)^3 }
    \sqrt{ \langle v_k^2 \rangle  }
  -
    \frac{ 3 }{2 (1+z)^2 }
    \frac{\langle  v_k \partial_z v_k \rangle}{  \sqrt{ \langle v_k^2 \rangle  }  }
  +
    \frac{1 }{4 (1+z) }
    \frac{ \langle  \partial_z v_k \partial_z v_k  \rangle   }{ \sqrt{ \langle v_k^2 \rangle  } }
  \right].
\label{eq:result_davt}
\end{align}

\vspace{2mm}
{\bf (2):} 
Since the $B$ coefficient has time dependence, it cannot
be easily factorized out from the second ensemble average
term in \geqn{eq:dphi_non_x2},
\begin{align} 
  \left\langle
    \frac{d (v_k A ) }{d t}
    \frac{ d [ v_k (B - A) ] }{d t}
  \right\rangle
=
  \left\langle
    \frac{d (v_k A ) }{d t}
    \frac{ d ( v_k B)  }{d t}
  \right\rangle
-
  \left\langle
    \frac{d (v_k A ) }{d t}
    \frac{ d ( v_k  A ) }{d t}
  \right\rangle.
\label{eq:dvkadtvkb}
\end{align} 
Note that the second term has already been derived
in \geqn{eq:result_davt}. Putting the time derivatives
$d (v_k A) / dt$ of \geqn{eq:dvkA_dt} and $d (v_k B) / dt$
of \geqn{eq:dvkb_dt} back into \geqn{eq:dvkadtvkb},
we can further simplify the first term,
\begin{align} 
  \left\langle
    \frac{d (v_k A ) }{d t}
    \frac{ d ( v_k B)  }{d t}
  \right\rangle  
& =
  - H^2
  \frac{\sqrt \pi}{3 k_{\rm fs}^3}
  \left( \frac{m_X}{T_{A 0}} \right)^{5/2} 
  \left[
    \frac{3 }{2 \sqrt{1+z}}
    \left\langle
      \frac{ v_k \partial_z v_k }{ \sqrt{ | v_k | } }
    \right\rangle
     B
  +
    \frac{3 }{2 \sqrt{1+z}}
    \left\langle
      \frac{ v_k^2 }{ \sqrt{ | v_k | } }
    \right\rangle
    \partial_z B
  \right.
\nonumber
\\
& \hspace{36mm}
\left.
  -
    \frac{\sqrt{1+z}}{2 }
    \left\langle
      \frac{ \partial_z v_k \partial_z v_k  }{ \sqrt{ |v_k| }} 
    \right\rangle
    B
  -
    \frac{\sqrt{1+z}}{2 }
    \left\langle
      \frac{ v_k \partial_z v_k  }{ \sqrt{ |v_k| }} 
    \right\rangle
    \partial_z B
  \right].
\label{eq:result_dabvt}
\end{align} 
While the second esemble average
$\langle v^2_k / \sqrt{|v_k|} \rangle$
can be replaced by $\langle v^2_k \rangle$
according to \geqn{eq:avg_x2_sqx}, the first
and the last ensemble averages in \geqn{eq:result_dabvt}
can be replaced by,
\begin{align} 
  \partial_z  \frac{v_k^2}{\sqrt{|v_k|}}
=
  \frac{ 2 v_k \partial_z v_k}{\sqrt{|v_k|}} 
- 
  \frac 1 2 
  v_k^2 
  \frac{ \partial_z |v_k| }{ |v_k|^{3/2} }
=
  \frac{ 2 v_k \partial_z v_k}{\sqrt{|v_k|}} 
- 
  \frac 1 2 
  v_k^2 
  \frac{v_k}{|v_k|}
  \frac{ \partial_z v_k }{ |v_k|^{3/2} }
=
  \frac 3 2
  \frac{ v_k \partial_z v_k}{\sqrt{|v_k|}},
\end{align}
in the similar way as \geqn{eq:pvkabsvk_new}.
Then, taking the ensemble average,
\begin{align} 
  \left\langle  \frac{ v_k \partial_z v_k}{\sqrt{|v_k|}} \right\rangle  
=
  \frac 2 3 \partial_z
  \left\langle  \frac{v_k^2}{\sqrt{|v_k|}} \right\rangle 
=
  \frac 2 3 
  \frac{ 2^{3/4}  \Gamma(5/4) }{ \sqrt{\pi} }
  \partial_z
  \langle v_k^2 \rangle^{3/4}
=
  \frac{ 2^{3/4}  \Gamma(5/4) }{ \sqrt{\pi} }
  \frac{ \langle v_k \partial_z v_k \rangle }
      {  \langle v_k^2 \rangle^{1/4} }. 
\label{eq:pvksqrtvk_new}
\end{align}

As already demonstrated in \geqn{eq:llfcvkr}, 
the time component can be factored out in a similar
way, such that the third term of \geqn{eq:result_dabvt}
becomes
\begin{align} 
  \left\langle
    \frac{ \partial_z v_k \partial_z v_k  }{ \sqrt{ |v_k| }} 
  \right\rangle  
=
  \frac{ 2^{3/4}  \Gamma(5/4) }{ \sqrt{\pi} }
  \frac{ \langle \partial_z v_k \partial_z v_k \rangle }
      {  \langle v_k^2 \rangle^{1/4} }, 
\label{eq:pzvkvk_sqvk}
\end{align} 
which can directly use \geqn{eq:coherent_velocity} now.

\vspace{2mm}
{\bf (3):}
The third ensemble average term \geqn{eq:dphi_non_x2} can be written as,
\begin{align} 
  \left\langle
    \frac{ d [ v_k (B - A) ] }{d t}
    \frac{ d [ v_k (B - A) ] }{d t}
  \right\rangle
=
  \left\langle
    \frac{ d ( v_k A ) }{d t}
    \frac{ d ( v_k A ) }{d t}
  \right\rangle
-
  2
  \left\langle
    \frac{ d ( v_k  A ) }{d t}
    \frac{ d ( v_k B  ) }{d t}
  \right\rangle
+
  \left\langle
    \frac{ d ( v_k B ) }{d t}
    \frac{ d ( v_k B ) }{d t}
  \right\rangle,
\end{align} 
where the first and second terms have already been
derived in \geqn{eq:result_davt} and \geqn{eq:result_dabvt}.
The third term using \geqn{eq:dvkb_dt},
\begin{align} 
  \left\langle
    \frac{ d ( v_k B ) }{d t}
    \frac{ d ( v_k B ) }{d t}
  \right\rangle
=
  (1+z)^2 H^2
  \left[
    \left\langle
      \partial_z v_k \partial_z v_k 
    \right\rangle
    B^2
  + 
    2  
    \left\langle
      v_k \partial_z v_k  
    \right\rangle
    B \partial_z B
  + 
    \left\langle
      v_k^2 
    \right\rangle
    (\partial_z B)^2
  \right],
\end{align}
where the esemble averages only involve
those terms already obtained in \geqn{eq:coherent_velocity}.

\vspace{3mm}
\noindent
{\bf (C-III) The variance $\langle \tilde \phi_X  \dot{\tilde \phi}_X \rangle $:}
Combining \geqn{eq:tilde_phiX} and \geqn{eq:tilde_dphiX}
to give
\begin{align} 
  \langle \tilde \phi_X  \dot{\tilde \phi}_X \rangle
& =
  \left( 4 \pi G F_X \rho_{\rm DM 0} \right)^2
  |\bm k|^2
\left\{ 
  \left\langle
    v_k A
    \frac{d (v_k A ) }{d t}
  \right\rangle
  \frac{  k_{\rm fs}^6 }{ \left( |\bm k| +  k_{\rm fs} \right)^6 }
+
  \left\langle
  v_k (B - A)
  \frac{ d [ v_k (B - A) ] }{d t}
  \right\rangle
  \frac{  k_{\rm fs}^8 }{ ( |\bm k| + k_{\rm fs})^8 }
\right.
\nonumber
\\
&
\hspace{32mm}
\left. +
\left[
  \left\langle
    v_k A
    \frac{ d [ v_k (B - A) ] }{d t}
  \right\rangle
+
  \left\langle
    v_k (B - A)
    \frac{d (v_k A ) }{d t}
  \right\rangle
\right]
  \frac{  k_{\rm fs}^7 }{ ( k + k_{\rm fs})^7 }
\right\}.
\label{eq:vkbaba}
\end{align} 

\vspace{2mm}
{\bf (1):}
Using \geqn{eq:phichi_vk_non} and \geqn{eq:dvkA_dt},
the first esemble average in \geqn{eq:vkbaba} can be written as,
\begin{align} 
  \left\langle
    v_k A
    \frac{d (v_k A ) }{d t}
  \right\rangle
= 
  \frac{\pi}{ 9 }
  \frac{1}{k_{\rm fs}^6}
  \left( \frac{m_X}{T_{A 0}} \right)^{5} 
  H
  \left[
    \frac{3 }{2 (1+z)^3 }
    \left\langle
    \frac{ v_k^2 }{  | v_k | }
    \right\rangle
  -
    \frac{1 }{2 (1+z)^2 }
    \left\langle
    \frac{ v_k \partial_z v_k }{ | v_k | }
    \right\rangle
  \right],
\label{eq:exp_vkadvka}
\end{align} 
where the ensemble averages $\langle v_k^2  / |v_k| \rangle $
and $ \langle v_k \partial_z v_k / |v_k| \rangle  $
are derived in \geqn{eq:avg_x2_absx} and \geqn{eq:pvkabsvk_new}.

\vspace{2mm}
{\bf (2):}
The final two ensemble average terms in \geqn{eq:vkbaba}
can be combined,
\begin{align} 
  \left\langle
    v_k A
    \frac{ d [ v_k (B - A) ] }{d t}
  \right\rangle  
+
  \left\langle
    v_k (B - A)
    \frac{d (v_k A ) }{d t}
  \right\rangle 
=
  \left\langle
    v_k A
    \frac{ d ( v_k B )  }{d t}
  \right\rangle  
-
  2
  \left\langle
    v_k A
    \frac{ d ( v_k  A)  }{d t}
  \right\rangle
+
  \left\langle
    v_k B
    \frac{d (v_k A ) }{d t}
  \right\rangle.
\label{eq:vkAdvbbma}
\end{align} 
Using \geqn{eq:phichi_vk_non} and \geqn{eq:dvkb_dt},
the first ensemble average can be written as,
\begin{align} 
  \left\langle
    v_k A
    \frac{ d ( v_k B )  }{d t}
  \right\rangle
=
  - H
    \frac{\sqrt \pi} 3
    \frac{1}{k_{\rm fs}^3}
    \left( \frac{m_X}{T_{A 0}} \right)^{5/2} 
    \frac 1 {\sqrt{1+z}}
  \left(
    \left\langle 
      \frac{ v_k  \partial_z v_k }{ \sqrt{ | v_k | } }
    \right\rangle 
    B
  +  
    \left\langle 
      \frac{ v_k^2 }{ \sqrt{ | v_k | } }  
    \right\rangle 
    \partial_z B  
  \right),
\label{eq:exp_vkadvkb}
\end{align} 
where $ \langle v_k \partial_z v_k / \sqrt{|v_k|} \rangle  $ and 
$\langle v_k^2 / \sqrt{|v_k|} \rangle $ can be found
in \geqn{eq:pvkabsvk_new} and \geqn{eq:avg_x2_sqx}.
The second term in \geqn{eq:vkAdvbbma} was already derived in \geqn{eq:exp_vkadvka}.
The last term in \geqn{eq:vkAdvbbma} can be written as,
using \geqn{eq:dvkA_dt}
\begin{align} 
  \left\langle
    v_k B
    \frac{d (v_k A ) }{d t}
  \right\rangle
=
    \frac{\sqrt \pi} 3
    \frac{1}{k_{\rm fs}^3}
    \left( \frac{m_X}{T_{A 0}} \right)^{5/2} 
    H
    \left[
      \frac{3 }{2 (1+z)^{3/2} }
      \left\langle
        \frac{ v_k^2 }{ \sqrt{ | v_k | } }
      \right\rangle
    -
      \frac{1 }{2 \sqrt{1+z}}
      \left\langle
        \frac{ v_k \partial_z v_k  }{ \sqrt{ |v_k| }} 
      \right\rangle
    \right]
  B,
\label{eq:exp_vkbdvkadt}
\end{align} 
where the first and second ensemble averages are given by
\geqn{eq:avg_x2_sqx} and \geqn{eq:pvksqrtvk_new}.

\vspace{2mm}
{\bf (3):} The second term in \geqn{eq:vkbaba} can
be written as,
\begin{align} 
  \left\langle
    v_k (B - A)
    \frac{ d [ v_k (B - A) ] }{d t}
  \right\rangle
& =
  \left\langle
    v_k B
    \frac{ d ( v_k B  ) }{d t}
  \right\rangle
-
  \left\langle
    v_k B
    \frac{ d ( v_k A ) }{d t}
  \right\rangle
-
  \left\langle
    v_k A
    \frac{ d (v_k B)}{d t}
  \right\rangle
+
  \left\langle
    v_k A
    \frac{ d (v_k  A)}{d t}
  \right\rangle,
\label{eq:vkbmadvmbma}
\end{align} 
where the first term is given by
\begin{align} 
  \left\langle
    v_k B
    \frac{ d ( v_k B  ) }{d t}
  \right\rangle
=
  - (1+z) H
  \left(
    \langle v_k \partial_z v_k \rangle B^2 
  + \langle v_k^2 \rangle  B \partial_z B
  \right),
\end{align} 
with the help of $d (v_k B) / d t$ in \geqn{eq:dvkb_dt}.
The second, third and last terms in \geqn{eq:vkbmadvmbma} are already calculated
in \geqn{eq:exp_vkbdvkadt}, \geqn{eq:exp_vkadvkb} and \geqn{eq:exp_vkadvka}, respectively.

\vspace{15mm}
\end{document}